# Validity of approximations applied in calculations of single-wall metallic carbon nanotube current-voltage characteristics

Dmitry Pozdnyakov

*Radiophysics and Computer Technologies Department of Belarusian State University, Nezavisimosty av. 4, 220030 Minsk, Belarus*

E-mail: pozdnyakov@tut.by; pozdnyakov@bsu.by

**Abstract** The calculation results of dependencies of electric current in the infinitely long single-wall metallic carbon nanotubes of armchair type with different diameter values on the strength of constant and uniform longitudinal electric field applied to them are presented in this study. On the one hand the results have been obtained by an accurate numerical self-consistent tight-binding full-band solution of the Boltzmann transport equations for electrons and phonons in the electric quantum limit. On the other hand the results have been obtained by solution of these equations after application of a number of commonly used simplifications and approximations. It is ascertained that the approximately calculated dependencies considerably diverge from the rigorously calculated ones. In addition the relaxation time of non-equilibrium phonons in the single-wall metallic carbon nanotubes has been estimated by means of comparison of the calculation results with the well-known experimental data. Its value is equal to $2.2 \pm 0.1$ ps.

## 1 Introduction

Presently the nanostructures with one-dimensional electron gas are very attractive objects of investigation for a large number of research laboratories and even whole research institutes. It is mainly caused by their advantages over the structures with two- and three-dimensional electron gas from practical point of view, particularly in the field of micro- and nanoelectronics. The carbon nanotubes, and specifically single-wall metallic carbon nanotubes (SWMCNTs), occupy a great niche among the structures like that [1–4]. SWMCNTs are studied quite well by now, but the issue connecting with an appropriate description of some their electrophysical properties still remains open. It has already been indicated in due course [1]. So, for example, several papers [5–10], in which the current-voltage characteristics (CVCs) of SWMCNTs were calculated with application of certain approximations for electron and phonon bandstuctures as well as for electron–phonon scattering rates, has been published by now. In one paper [5] the calculations were realized by a numerical solution of the Boltzmann transport equation for electrons. In three other papers [6–8] the Boltzmann transport equations for electrons and phonons were self-consistently solved by numerical methods. In two other papers [9, 10] CVCs of nanotubes were calculated by a numerical solution of both the Boltzmann transport equation for electrons and the thermal conductivity equation. Moreover, to calculate CVCs of SWMCNTs other methods were applied in a number of papers. Namely the calculations were based on Landauer-Buttiker formalism [11, 12], Monte Carlo method [13] and others [14]. But, in spite of an impressive number of publications, the problem of appropriate description of SWMCNT electrophysical properties still remains unsolved. And here are the reasons. First, different phonon modes were either taken into account or not in the calculations [5–10]. Second, such fitting parameter as the relaxation time (mean lifetime) of the non-equilibrium phonons in the nanotube [6–10] $\tau$ was associated with either the phonon–phonon scattering processes ($\tau \sim 1..10$ ps) [6–8] or the direct escape of the non-equilibrium phonons from the nanotube into a substrate ($\tau \sim 1..100$ fs) [8–10] owing to the immediate interaction of the nanotube atoms with the substrate atoms [11, 12]. Moreover, from paper to paper the values of the fitting parameter $\tau$ greatly varied from 6.9 fs [9] to 5.3 ps [6] for SWMCNTs on the insulating substrates. This is very troubling since in the different publications the calculated CVCs were frequently fitted with $\tau$ to the same experimental characteristics. At that the calculated effective temperature of the non-equilibrium phonon gas was sometimes considerably larger than the melting or even the boiling temperature of graphite (graphene) that seems to be unbelievable.

Taking into account all the mentioned above, it is possible to formulate the tasks which being solved will allow the processes taking place in SWMCNTs and characterizing their electrophysical properties to be understood correctly and described appropriately. In turn, it will allow CVCs of such nanotubes to be calculated rigorously. In particular, some of such important tasks are the following: (a) calculation of CVCs of the considered nanotubes in the framework of rigorous quantum-mechanical description of electron–phonon interaction; (b) determination of a real value of parameter $\tau$ by comparison of calculated results with the experimental data; (c) ascertainment of a deviation degree of SWMCNT CVCs obtained with application of commonly used approximations from the rigorously calculated ones.

Thus the purpose of this article is a solution of the three aforesaid tasks.

## 2 Theory

### 2.1 Electron–phonon scattering

Without loss of generality let us consider such SWMCNTs as $(N, N)$ nanotubes of "armchair" type with a chiral index $N \leq 18$ which are on some insulating substrate. The nanotube diameter limitation allows the electric quantum limit for electrons and phonons in the nanotubes to be applied [5–17] as, in spite of considerable heating of electron and phonon gases in high electric fields [6–12, 14, 18], in principle, only the nearest to the Fermi level electron bands [15, 16, 19] and the zero angular momentum phonon bands [15, 16, 20] can be taken into consideration for SWMCNTs with the diameter value equal to 2.44 nm or less [18]. Let us also consider the approximation of an ideal thermal contact between the nanotube and a substrate. In our case this term means that the nanotube long-wave acoustic phonons are always in thermodynamic equilibrium with the substrate long-wave acoustic phonons [6, 7].

It should be noted that foundations of the theory of electron–phonon scattering in SWMCNTs of "armchair" type (SWACNTs) has been formerly developed [19] in the framework of both the tight-binding approximation and approximation of the tight-binding Hamiltonian linear perturbation by the phonon assisted displacements of atoms from their equilibrium positions in the crystal lattice [21, 22]. The starting point of the theory is consideration of the vibrations of carbon atoms as the plane phonon modes frozen into graphene crystal lattice [20]. Such a description of phonons in SWACNTs along with the application of the electric quantum limit approximation [19, 20] allows both *ab initio* quantum-mechanical calculations for electron–phonon coupling to be carried out and expressions for calculation of electron–phonon scattering rates to be rigorously obtained in an analytical form [19, 20]. Let us write down the formulae allowing the scattering rates of electrons by the phonons frozen into the lattice of SWACNT to be calculated with the assumption that all the final electron states are free (the occupation of the final states will be directly taken into account by means of introduction of certain multipliers into the scattering operators). They are

$$W_{\text{LA}}^{\text{e/a}}(k) = w_N \frac{3J_0}{\hbar\omega_{\text{LA}}^{\text{e/a}}} \left|\sin\left(\frac{ak_{\text{LA}}^{\text{e/a}}}{2}\right)\right|^{-1} \left(n_{\text{LA}}(q_{\text{LA}}^{\text{e/a}}) + \frac{1}{2} \pm \frac{1}{2}\right) \sin^2\left(a(k_{\text{LA}}^{\text{e/a}} - k)/4\right) \cos^2\left(a(k_{\text{LA}}^{\text{e/a}} + k)/4\right), \quad (1)$$

$$W_{\text{TA}}^{\text{e/a}}(k) = w_N \frac{J_0}{\hbar\omega_{\text{TA}}^{\text{e/a}}} \left|\sin\left(\frac{ak_{\text{TA}}^{\text{e/a}}}{2}\right)\right|^{-1} \left(n_{\text{TA}}(q_{\text{TA}}^{\text{e/a}}) + \frac{1}{2} \pm \frac{1}{2}\right) \sin^2\left(a(k_{\text{TA}}^{\text{e/a}} - k)/4\right) \sin^2\left(a(k_{\text{TA}}^{\text{e/a}} + k)/4\right), \quad (2)$$

$$W_{\text{LO}}^{\text{e/a}}(k) = w_N \frac{3J_0}{\hbar\omega_{\text{LO}}^{\text{e/a}}} \left|\sin\left(\frac{ak_{\text{LO}}^{\text{e/a}}}{2}\right)\right|^{-1} \left(n_{\text{LO}}(q_{\text{LO}}^{\text{e/a}}) + \frac{1}{2} \pm \frac{1}{2}\right) \cos^2\left(a(k_{\text{LO}}^{\text{e/a}} - k)/4\right) \sin^2\left(a(k_{\text{LO}}^{\text{e/a}} + k)/4\right), \quad (3)$$

$$W_{\text{TO}}^{\text{e/a}}(k) = w_N \frac{J_0}{\hbar\omega_{\text{TO}}^{\text{e/a}}} \left|\sin\left(\frac{ak_{\text{TO}}^{\text{e/a}}}{2}\right)\right|^{-1} \left(n_{\text{TO}}(q_{\text{TO}}^{\text{e/a}}) + \frac{1}{2} \pm \frac{1}{2}\right) \left\{1 + \cos\left(a(k_{\text{TO}}^{\text{e/a}} - k)/4\right)\cos\left(a(k_{\text{TO}}^{\text{e/a}} + k)/4\right)\right\}^2, \quad (4)$$

$$w_N = \frac{1}{N} \frac{\hbar \alpha_0^2}{2m_C}, \quad (5)$$

$$k_{\text{LA,TA,LO,TO}}^{\text{e/a}} = q_{\text{LA,TA,LO,TO}}^{\text{e/a}} + k, \quad (6)$$

$$E_{1,2}(k_{\text{LA,TA,LO,TO}}^{\text{e}}) = E_{1,2}(k) - \hbar\omega_{\text{LA,TA,LO,TO}}^{\text{e}}(q_{\text{LA,TA,LO,TO}}^{\text{e}}), \quad (7)$$

$$E_{1,2}(k_{\text{LA,TA,LO,TO}}^{\text{a}}) = E_{1,2}(k) + \hbar\omega_{\text{LA,TA,LO,TO}}^{\text{a}}(q_{\text{LA,TA,LO,TO}}^{\text{a}}), \quad (8)$$

$$E_{1,2}(k) = \mp J_0 \left[1 - 2\cos(ak/2)\right]. \quad (9)$$

Here $W$ is the scattering rate, $\hbar$ is the Plank constant, $m_C$ is the carbon atom mass, $a = 2.46$ Å is the graphene lattice spacing [20, 23, 24], $\alpha_0 = 2.16$ Å$^{-1}$ is the deformation constant for graphene [20, 22, 25], $J_0 = 2.7$ eV is the graphene $\pi$-orbital hopping parameter [20, 26], $q$ is the phonon wave vector ($q \in [-2\pi/a, 2\pi/a]$), $\omega$ is the cyclic phonon frequency, $n$ is the phonon population number (here and further it is the same that the phonon distribution function [7]), $k$ is the electron wave vector ($k \in [-\pi/a, \pi/a]$), $E_1$ and $E_2$ is the electron energy in the first and the second band, respectively, relative to the Fermi level which is taken as zero ($E_F = 0$) [19]. The superscripts "e" and "a" denote the phonon emission and absorption processes, respectively. The subscripts "LA", "LO", "TA" and "TO" denote [20, 27] the longitudinal acoustic, longitudinal optic, in-plane transverse acoustic and in-plane transverse optic phonon modes, respectively. The processes of electron scattering by LA and TO phonons are intraband (band 1 ↔ band 1 and band 2 ↔ band 2), and the processes of electron scattering by TA and LO phonons are interband (band 1 ↔ band 2) [1, 19, 20]. Equations (1) – (4) was strictly derived basing on Pietronero–Strassler–Zeller–Rice calculations [22] with application of a number of results from refs. [19, 20]. Secondary quantum effects showing appreciably their worth just in the quantum systems with the one-dimensional electron gas [28] were not taken into account during the derivation since the ranges of electron energy, where they could considerably influence the electron scattering processes in SWACNTs, are sufficiently far from the Fermi level.

Regarding eqs. (1) – (8) it is necessary to make a few remarks and explanations. There are no expressions for calculation of scattering rates of electrons by the out-of-plane transverse (radial breathing) acoustic (ZA in [27], and RBM in [1]) and out-of-plane transverse optic (ZO in [27], and oTO in [1]) phonon modes among them as such phonon modes do not cause a linear perturbation of the tight-binding Hamiltonian by carbon atom displacements [19, 20, 22]. Using these formulae, it is enough to calculate the electron scattering rates only for the cases when the electron group velocity vector reverses (the backward scattering), and it is no need to take into consideration the forward scattering. The wave-vector $k$ reverses after the backward scattering of an electron by LA and TO phonons, and it does not reverse after the backward scattering of an electron by LO and TA phonons [19, 20]. It should be noted that in a number of papers the phonon modes causing the backward scattering of electrons are often called and denoted in a different way. Namely, the short-wave LA phonons are called as zone-boundary $E_1'$ phonons [1, 30], the short-wave TO phonons are called as zone-boundary $A_1'$ (or K) phonons [1, 5–7, 9, 17, 23, 24, 30], the long-wave LO phonons are called as $E_{2g}$ (or Γ) phonons [5–7, 9, 17, 23, 24, 30], and the long-wave TA phonons are called as twisting phonons, twistons (TW) or simply acoustic phonons [1, 5–7, 17]. The fact is that the forward scattering of active electrons (the charge carriers with energies close to the Fermi level) [15] can be neglected in view of the following reasons: the state of an electron is not changed for its forward scattering by LA phonon; there are differences of several orders of magnitude between the forward scattering rates of electrons by LO and TA phonons and the backward scattering rates of charge carriers by LA, LO or TO phonons; the contribution of forward scattering of electrons by TO phonons to the non-equilibrium electron distribution function relaxation is either minor in comparison with the other scattering processes in case of not too large difference between the effective temperature of electron gas [9, 10] $T_e$ and the effective temperature of TO phonon gas for the phonons with wave vector values in the vicinity of $\omega_{TO}(0)/u_F$ ($u_F = \sqrt{3}aJ_0/2\hbar$ is the absolute value of electron group velocity at the Fermi level [15]), or vanishingly small in the case of equality of these temperatures. Moreover the forward TO phonon scattering of electrons is responsible only for the relaxation of non-equilibrium electron gas temperature $T_e$ to the nanotube temperature. It almost does not take part in the relaxation of quasi-Fermi levels.

Eqs. (1) – (4) have been obtained in the electric quantum limit approximation which was explicitly or implicitly applied in all the papers (see refs. [1, 5–17, 19, 23, 31]) excepting ref. [18]. Such an approximation appropriately describes the scattering processes with participation of both the electrons belonging to the first or the second band of the SWACNT first electron Brillouin zone [15, 16, 18–20] and the phonons with zero angular momentum [15, 16, 20] being the coherent lattice vibrations [29]. The bandstructure for such phonons is represented in fig. 1.

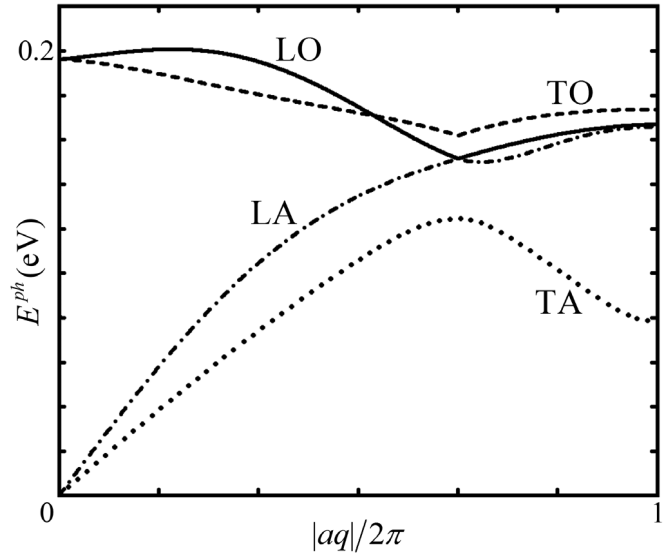

**Fig. 1** The phonon bandstructure of SWACNTs for the phonons with zero angular momentum

The dependencies shown in fig. 1 have been calculated in the framework of approach proposed in refs. [15, 16]. It is based on the analytic representation of *ab initio* calculated phonon dispersion curves by means of polynomials and harmonic functions. The approach like that allows eqs. (6) – (8) to be quickly solved by numerical methods with high accuracy. In particular, these dependencies have been obtained by means of zone-folding method conformably to the phonon bandstructure of graphene from refs. [24, 27] taking into account Kohn anomalies [24]. As is well known the curvature effects are neglected in that case [23, 30]. However, the validity and very good accuracy of zone-folding approximation for the phonon bandstructure of SWACNTs have already been proved in refs. [23, 30]. The fact that the first Brillouin zone boundary for phonons lies at a distance of $2\pi/a$ from the zone centre but not $\pi/a$ is caused by application of zone-folding method conformably to the modified phonon bandstructure of graphene [16]. The consideration of modified phonon bandstructure of graphene allows both the normal and umklapp scattering processes to be reduced only to the normal processes [16, 20].

As an example the backward scattering rates of electrons by the phonons in (15,15) SWACNT are presented in fig. 2 for a case of the nanotube equilibrium (substrate) temperature $T = 290$ K.

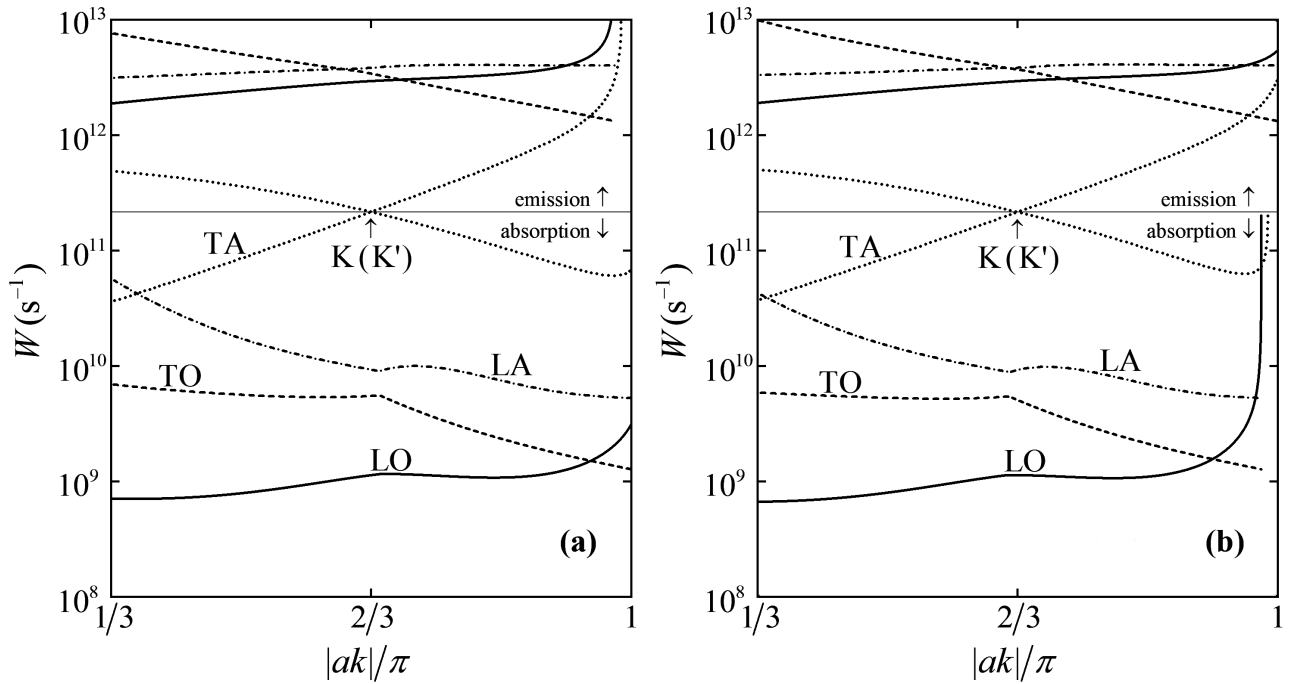

**Fig. 2** The backward electron–phonon scattering rates in (15,15) SWACNT:
(a) – scattering rates for the first electron band, (b) – scattering rates for the second one

## 2.2 Electron–phonon transport

Now let us write down the system of Boltzmann transport equations for electrons and phonons which being solved self-consistently allows the value of any electrophysical parameter and any electrical characteristic of SWACNTs to be calculated. The system can be represented like this [6–9, 15, 31]

$$\begin{cases} \partial_t f_{1,2} = e\hbar^{-1} F(x,t) \cdot \partial_k f_{1,2} \pm a J_0 \hbar^{-1} \sin(ak/2) \cdot \partial_x f_{1,2} - \mathrm{Sc}_f[f_{1,2}], \\ \partial_t n_{\mathrm{LA,LO,TO}} = \dfrac{n^0_{\mathrm{LA,LO,TO}} - n_{\mathrm{LA,LO,TO}}}{\tau_{\mathrm{LA,LO,TO}}(x,q,\ldots)} - \partial_q \omega_{\mathrm{LA,LO,TO}} \cdot \partial_x n_{\mathrm{LA,LO,TO}} - \mathrm{Sc}_n[n_{\mathrm{LA,LO,TO}}]. \end{cases} \quad (10)$$

Here $e$ is the value of elementary charge, $F$ is the longitudinal electric field strength in the nanotube, $f$ is the electron distribution function for the corresponding band, $n^0 = n^0(T)$ is the equilibrium phonon (Bose–Einstein) distribution function. In the second equation of the system (10) the phonon–phonon scattering operator (the first term after sign of equality) is written down in the approximation of relaxation time [6–8]. The electron scattering operator $\mathrm{Sc}_f$ effects on $f_i$ in such a way (the occupation of the final states is taken into account) [6, 7, 9, 15]

$$\mathrm{Sc}_f[f_i(k)] = \sum_{j,\,k'=\{k_j^e,k_j^a\}} \left[ W^{e/a}_{j,i\to i'}(k) f_i(k)(1-f_{i'}(k')) - W^{a/e}_{j,i'\to i}(k') f_{i'}(k')(1-f_i(k)) \right], \quad (11)$$

where $i = \{1,2\}$, $j = \{\mathrm{LA, LO, TA, TO}\}$. The phonon scattering operator $\mathrm{Sc}_n$ effects on $n_{\mathrm{LA}}$ and $n_{\mathrm{TO}}$ like this

$$\mathrm{Sc}_n[n(q)] = 2\left|\frac{dk_1}{dq}\right| \left\{ W^a_{1\to 1}(k_1) f_1(k_1)(1-f_1(k_1^a)) - W^e_{1\to 1}(k_1^a) f_1(-k_1^a)(1-f_1(-k_1)) \right\} +$$
$$2\left|\frac{dk_2}{dq}\right| \left\{ W^a_{2\to 2}(k_2^e) f_2(-k_2^e)(1-f_2(-k_2)) - W^e_{2\to 2}(k_2) f_2(k_2)(1-f_2(k_2^e)) \right\}, \quad (12)$$

and it effects on $n_{\mathrm{LO}}$ in such a way

$$\mathrm{Sc}_n[n(q)] = 2\left|\frac{dk_1}{dq}\right| \left\{ W^a_{1\to 2}(k_1) f_1(k_1)(1-f_2(k_2^a)) - W^e_{2\to 1}(k_1) f_2(k_1)(1-f_1(k_1^e)) + \right.$$
$$\left. W^a_{1\to 2}(k_1^e) f_1(-k_1^e)(1-f_2(-k_1)) - W^e_{2\to 1}(k_2^a) f_2(-k_2^a)(1-f_1(-k_1)) \right\} +$$
$$2\left|\frac{dk_2}{dq}\right| \left\{ W^a_{2\to 1}(k_2) f_2(k_2)(1-f_1(k_1^a)) - W^e_{1\to 2}(k_2) f_1(k_2)(1-f_2(k_2^e)) + \right.$$
$$\left. W^a_{2\to 1}(k_2^e) f_2(-k_2^e)(1-f_1(-k_2)) - W^e_{1\to 2}(k_1^a) f_1(-k_1^a)(1-f_2(-k_2)) \right\}. \quad (13)$$

The arguments $x$ and $t$ of functions $f$ and $n$ are omitted in eqs. (11) – (13) for brevity. In eqs. (12) and (13) the relations between the initial $k_{1,2}$ and final $k_{1,2}^{e/a}$ wave vectors for an electron and the wave vector $q$ for a phonon are determined by the energy and momentum conversation laws according to eqs. (6) – (8). There is a value $2|dk/dq|$ in these equations because of the need to renormalize the density of electron states to the density of phonon states. The fact is that the right parts of these equalities are represented through the electron but not phonon states. It is more convenient for a numerical solution of the system (10). It should be noted that the value $2|dk/dq|$ slightly differs from unity for all the scattering mechanisms of electrons in the wide range of their energy in the vicinity of the Fermi level. As a result this multiplier can be omitted, and eqs. (12) and (13) can be reduced to the equations similar to the expressions for the phonon scattering operators from refs. [6, 7].

The Boltzmann transport equation for TA phonons does not enter the system (10) because according to the other papers, in which TA phonon scattering of electrons is taken into consideration, it is assumed that such a phonon gas is in thermodynamic equilibrium with the substrate phonon gas [6, 7, 12, 13, 15, 17, 31], that is $n_{\mathrm{TA}} = n^0_{\mathrm{TA}}$.

# 3 Results of calculations and discussion

## 3.1 Rigorous calculations

As the experimental results show [13, 17, 32] SWMCNT CVCs measured for the same values of both the nanotube diameter $d$ and the temperature of the substrate and electrodes $T$ but for different values of the nanotube length $L$ are scaled very well. That is, the dependencies of the electric current $I$ in SWMCNTs on the longitudinal electric field strength $F_0 = V/L$, where $V$ is the voltage applied to SWMCNT, differ from each other insignificantly for $L > 0.1$ μm. And this is despite both the various kinds of metals, of which the electrodes guaranteeing the good ohmic contacts are made, and various kinds of materials of which the heat-eliminating insulating substrates are made. Well, if CVCs have been measured for the same nanotube, for example, by means of the scanning atomic force microscopy methods, then the scaled CVCs almost completely coincide with each other [17]. Taking into account these facts as well as necessity to ascertain SWACNT CVC main features characterizing the nanotubes themselves but not the nanotube–metal contacts, let us resort to an approximation often used for simplification of the Boltzmann equation solution. Let us assume that the first and second terms of kinetic equations entering the system (10) are identically equal to zero and $F(x, t) = F_0 = const$. From the physical point of view the approximation like that is equivalent to consideration of the steady-state transport of electrons and phonons in an infinitely long SWACNT. Such a transport is a limiting case of electron–phonon transport in a rather long SWACNT on an insulating substrate with ideal heat conducting, an ideal thermal contact also takes place, and between electrodes with ideal electrical conducting, an ideal ohmic contact takes place too [15]. At that it should be noted that the conditions of electron transport in SWACNTs being on titanium electrodes are, in principle [2, 33], very similar to the conditions of steady-state transport of charge carriers in infinitely long SWACNTs when $L > 2l_{max} = 2v_F \max\{\tau_0, 1/W_{TA}^{e+a}(k_F)\} \approx 4$ μm at $T \geq 290$ K. Besides, by analogy with the other papers (see, for example, [6–10, 31]) let us consider an effective relaxation time for all the non-equilibrium phonons $\tau_0$ such that $\tau_{LA,LO,TO}(x, q, \ldots) \approx \tau_0 = const$.

The calculation results for the dependencies of electric current $I$ in SWACNTs at $T = 290$ K, $\tau_0 = 2.2$ ps and the different values of their diameter $d = Na\sqrt{3}/\pi$ on the strength $F_0$ of the longitudinal electric field applied to the nanotubes are presented in fig. 3. The results have been obtained after the direct numerical self-consistent solution of the five Boltzmann transport equations (two for electrons in the different bands and three for LA, LO and TO phonons) entering the system (10) using a finite-difference method.

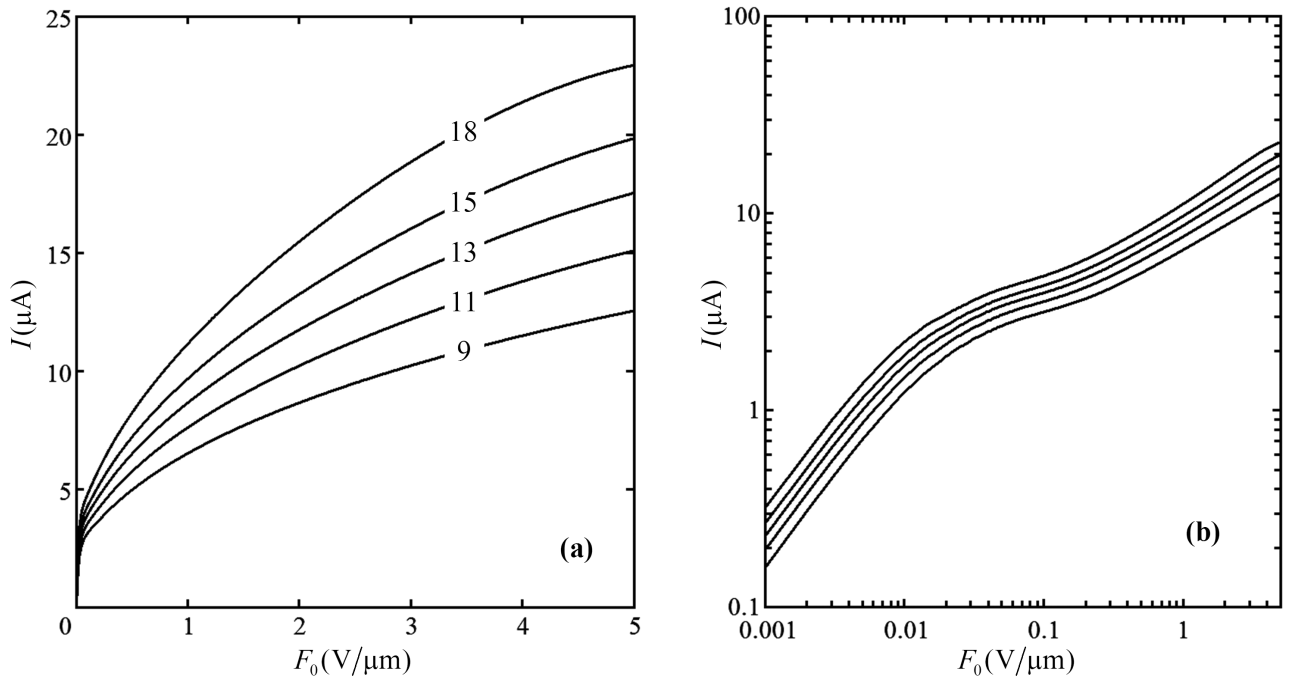

**Fig. 3** The exactly calculated dependencies of $I$ on $F_0$ for the infinitely long $(N, N)$ SWACNTs: (a) – linear scale, (b) – logarithmic scale

The dependencies shown in fig. 3 have been numerically calculated by formula

$$I(t) = \frac{eaJ_0}{\pi\hbar} \int_{-\pi/a}^{+\pi/a} (f_1(k,t) - f_2(k,t)) \sin(ak/2) dk \quad (14)$$

in accordance with the expressions from ref. [15] which allow the electric current in SWACNTs to be calculated rigorously.

**3.2 Approximate calculations**

Along with the rigorously calculated curves the analogous dependencies obtained after the application of a number of commonly used simplifications and approximations (see, for example, [5–14, 17, 23]) are represented in fig. 4.

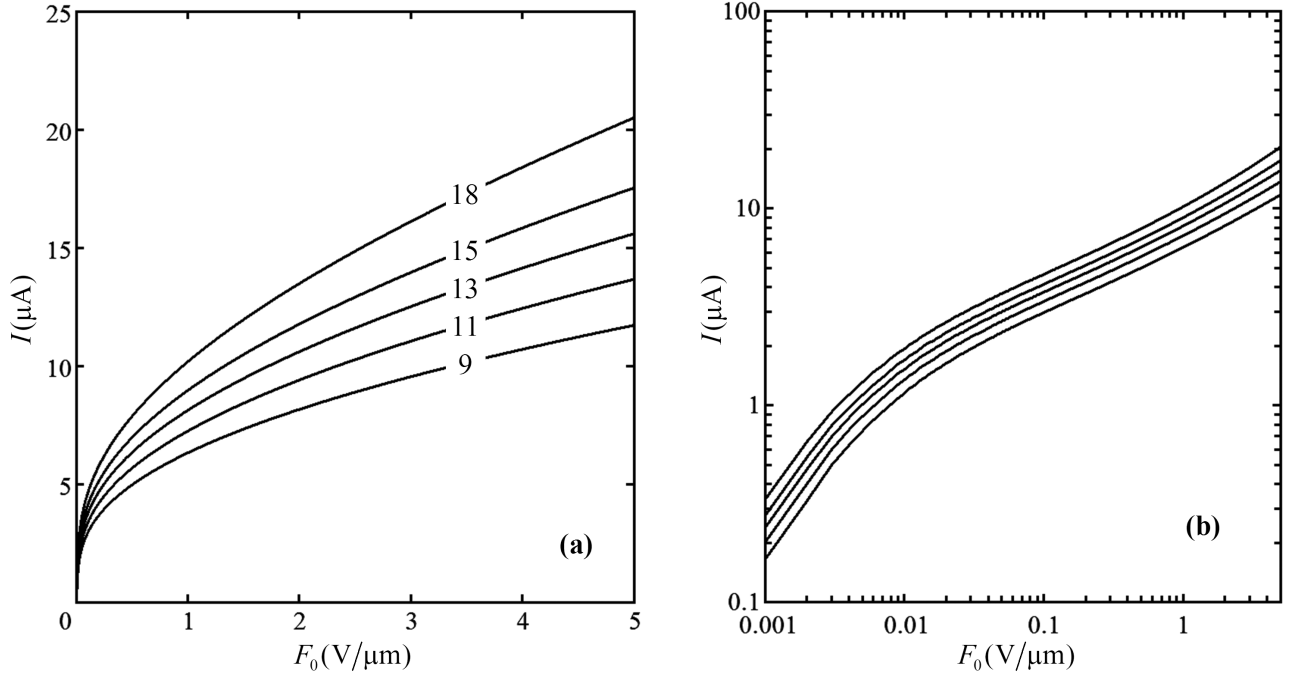

**Fig. 4** The approximately calculated dependencies of $I$ on $F_0$ for the infinitely long $(N, N)$ SWACNTs: (a) – linear scale, (b) – logarithmic scale

In particular, the following simplifications and approximations have been applied. It is the linear dispersion relation approximation (LDRA) for electrons (used in refs. [5–14, 17, 23]), i. e. $E_{1,2} = \mp u_F \hbar K_{1,2}$ and $E_{1,2} = \pm u_F \hbar K'_{1,2}$. $K$ and $K'$ are the electron wave vectors measured respectively from the Dirac points K and K' in graphene (nanotube). It is the symmetric function behaviour approximation (SFBA) in the framework of which it is supposed that the behaviour of functions $f_1$ and $f_2$ is symmetric in the vicinity of K and K' points. As a result it is possible to consider only one of these points, and the other can be taken into account in the calculations by introducing the degeneracy factor $g_{KK'} = 2$ (used explicitly or implicitly in refs. [5–14, 17]). It is the constant scattering rate approximation (CSRA) for scattering of the active electrons by the phonons (used in refs. [5–14, 17, 23]). It is the elastic electron scattering approximation (EESA) for scattering of electrons by TA phonons (used in refs. [6, 7, 11–13, 17]). LDRA, CSRA and EESA allow the electron scattering rate calculation to be simplified very much. As a result eqs. (1) – (4) can be reduced to the approximate equalities

$$W_{LA,LO,TO}^{e/a} \approx \frac{1}{N} \frac{3u_F \hbar^2 \alpha_0^2}{2m_C a E_{LA,LO,TO}^{ph}} \left( n(E_{LA,LO,TO}^{ph}) + \frac{1}{2} \pm \frac{1}{2} \right) =$$

$$\frac{u_F}{l_{LA,LO,TO}} \left( n(E_{LA,LO,TO}^{ph}) + \frac{1}{2} \pm \frac{1}{2} \right) = \frac{u_F}{\eta_{LA,LO,TO} d} \left( n(E_{LA,LO,TO}^{ph}) + \frac{1}{2} \pm \frac{1}{2} \right), \quad (15)$$

$$W_{el} = W_{el}^e + W_{el}^a \approx \frac{1}{N} \frac{u_F \alpha_0^2 a k_B T}{16 m_C u_{TA}^2} = \frac{u_F}{l_{TA}} \frac{T}{T_0} = \frac{u_F}{\eta_{TA} d} \frac{T}{T_0}. \quad (16)$$

Here $E_{LA}^{ph} = 151.4\,\text{meV}$, $E_{LO}^{ph} = 196.1\,\text{meV}$, $E_{TO}^{ph} = 161.8\,\text{meV}$, $u_{TA} = 13.92\,\text{km/s}$ [24, 27], $\eta_{LA} = 112.5$, $\eta_{LO} = 146.7$, $\eta_{TO} = 121.5$, $\eta_{TA} = 961.7$, $T_0 = 300$ K, $k_B$ is the Boltzmann constant. And finally, it is the hot electrons and quasi-Fermi levels approximation (HEQFLA) which has been applied in refs. [9, 10]. It is based on consideration of the Fermi–Dirac functions, that are characterized by the quasi-Fermi levels (shifted Fermi levels) and the temperature of hot electron gas $T_e$ greater than the substrate temperature $T$, instead of the non-equilibrium electron distribution functions. In the framework of the latter approximation the equations

$$\pm u_F e F_0 \partial_{E_{1,2}} f_{1,2} + \text{Sc}_f [f_{1,2}] = 0 \quad (17)$$

obtained from the couple of differential equations of the system (10) after application of LDRA (see, for example, [5–9]) and solved in the vicinity of K point are reduced after application of SFBA, CSRA, EESA, eqs. (15) and (16) to the couple of transcendental equations with the desired values of parameters $T_e$, $\Delta E_{F1}$ and $\Delta E_{F2}$ ($\Delta E_{F1} = -\Delta E_{F2}$) in the points $E_{1,2} = \pm \Delta E_{F1}$ for $f_1$ or in the points $E_{1,2} = \pm \Delta E_{F2}$ for $f_2$ ($f_1(E) = 1 - f_2(-E)$).

### 3.3 Discussion of the calculation results

It follows from figs. 3 and 4 that SWMCNT CVCs must directly depend on the nanotube diameter since the value of SWMCNT resistance per unit length $\rho = I/F_0$ decreases with increase of the diameter value $d$. It is obvious that such a dependence is a result of relations $\rho \sim W \sim 1/d \Leftrightarrow I \sim d = Na\sqrt{3}/\pi$. The same conclusion has been made in refs. [8, 9] for rather long nanotubes. The comparison of data in figs. 3 and 4 shows that the value of relative difference $(I_{ex} - I_{ap})/I_{ap}$ between the rigorous ($I_{ex}$) and approximate ($I_{ap}$) calculations of electric current $I$ in SWACNTs is more than 0.1 for $F_0$ from 0.15 to 0.6 V/µm and more than 0.05 for $F_0$ from 1.5 to 5 V/µm. Moreover, as the analysis of a whole number of additional calculation results showed, the maximal dominant error in the calculations is introduced by HEQFLA, and the minimal error is introduced by EESA and LDRA. As a result, an estimate of SWMCNT CVCs can be made applying CSRA, EESA, SFBA and LDRA. More rigorous calculations of CVCs with a quite acceptable degree of precision can be carried out with application of only EESA and LDRA.

It is necessary to note that the value of parameter $\tau_0 = 2.2$ ps has not been chosen randomly. The fact is that the calculated value of SWACNT differential resistance per unit length $\delta\rho(F_0) = (dI/dF_0)^{-1}$ of (13,13) ($d = 1.8$ nm) and (15,15) ($d = 2.0$ nm) SWACNTs is equal to 0.78 MOhm/µm at $F_0 = 5$ V/µm and $T = 290$ K only if the value of parameter $\tau_0$ is equal to 2.2 ps. In this case it coincides with the value of $\delta\rho_{high} = 0.78 \pm 0.04$ MOhm/µm experimentally measured by J.-Y. Park, S. Rosenblatt et al. [17] (the same value of $\delta\rho_{high}$ one can calculate from the experimentally measured SWACNT CVCs represented in refs. [5, 13]). As the simulation results show the value of $\delta\rho$ strongly depends on $\tau_0$ at $F_0 \gg 0.1$ V/µm (see, for example, [6]). The variation of parameter $\tau_0$ in the range $\pm 0.1$ ps relative to the value 2.2 ps causes the variation of the value $\delta\rho$ in the range $\pm 0.04$ MOhm/µm relative to the value 0.78 MOhm/µm for (13,13) and (15,15) SWACNTs at $F_0 = 5$ V/µm and $T = 290$ K. The calculation results along with the experimental data of refs. [5, 13, 17] allow the value of $\tau_0$ to be determined. It is equal to $2.2 \pm 0.1$ ps. Such a value of parameter $\tau_0$ for SWACNTs agrees with both the various experimental measurements for graphene $\tau_{LO} = 2.2 \pm 0.1$ ps [34], $\tau_0 = 2.1 \pm 0.4$ ps [35] ($\tau_0 = 2.8 \pm 0.4$ ps [35] and $\tau_0 = 2.55 \pm 0.1$ ps [36] for double-layer graphite) and *ab initio* theoretical calculations for this matter [37] $\tau_{LA} = 1.0$–1.6 ps, $\tau_{LO} = 2.0$–3.2 ps, $\tau_{TO} = 2.3$–4.8 ps. And this clearly indicates ($\tau_0 \in (1, 10)$ ps and $\tau_0 \notin (1, 100)$ fs) that the dominant mechanism of relaxation of the non-equilibrium phonons in the nanotubes is always a phonon–phonon scattering regardless of whether the nanotubes contact with a substrate or not (the value of $\tau_0$ is the same for supported and suspended graphene [35]). In addition, it should be noted that the calculated value of SWACNT differential resistance per unit length $\delta\rho(F_0)$ for (13,13) nanotube is equal to 4.3 kOhm/µm at $F_0 = 0$ and $T = 290$ K. It also coincides with the experimentally measured value $\delta\rho_{low} = 4.3 \pm 0.3$ kOhm/µm [17].

### 3.4 Conclusion

The calculation of CVCs of infinitely long SWACNTs has been carried out in the framework of rigorous full-band quantum-mechanical description of electron–phonon interaction in the electric quantum limit. A deviation degree of CVCs of infinitely long SWACNTs obtained with application of commonly used approximations from the rigorously calculated ones has been ascertained. In particular, less rigorous calculations of CVCs with still a quite acceptable degree of precision can be carried out with application of only EESA and LDRA. One can only estimate SWMCNT CVCs while applying such approximations as CSRA, EESA, SFBA and LDRA. The application of HEQFLA along with CSRA, EESA, SFBA and LDRA results in wrong calculations. In addition the real value of parameter $\tau$ has been determined by comparison of the calculated results with the known experimental data. Its value is equal to $2.2 \pm 0.1$ ps. It has been also ascertained that the dominant mechanism of relaxation of the non-equilibrium phonons in the nanotubes is always a phonon–phonon scattering regardless of whether the nanotubes contact with some substrate or not.

## Acknowledgement

The author is very grateful to Springer for publication of the manuscript in "Journal of Computational Electronics". The final publication is available at http://link.springer.com/. DOI 10.1007/s10825-012-0419-6
## References

1. Roche, S., Jiang, J., Torres, L. E F F., Saito, R.: Charge transport in carbon nanotubes: Quantum effects of electron–phonon coupling. J. Phys.: Condens. Matter **19**, 183203 (2007).
2. Lim, S.Ch., Jang, J.H., Bae, D.J., Han, G.H., Lee, S., Yeo, I.-S., Lee, Y.H.: Contact resistance between metal and carbon nanotube interconnects: Effect of work function and wettability. Appl. Phys. Lett. **95**, 264103 (2009).
3. Xie, L., Farhat, H., Son, H., Zhang, J., Dresselhaus, M.S., Kong, J., Liu, Z.: Electroluminescence from suspended and on-substrate metallic single-walled carbon nanotubes. Nano Lett. **9**, 1747 (2009).
4. Topinka, M.A., Rowell, M.W., Goldhaber-Gordon, D., McGehee, M.D., Hecht, D.S., Gruner, G.: Charge transport in interpenetrating networks of semiconducting and metallic carbon Nanotubes. Nano Lett. **9**, 1866 (2009).
5. Yao, Z., Kane, Ch.L., Dekker, C.: High-field electrical transport in single-wall carbon nanotubes. Phys. Rev. Lett. **84**, 2941 (2000).
6. Lazzeri, M., Mauri, F.: Coupled dynamics of electrons and phonons in metallic nanotubes: Current saturation from hot-phonon generation. Phys. Rev. B **73**, 165419 (2006).
7. Auer, C., Schurrer, F., Ertler, C.: Influence of hot phonons on the transport properties of single-wall carbon nanotubes. J. Comput. Electron. **6**, 325 (2007).
8. Hasan, S., Alam, A., Lundstrom, M.: The role of self-heating and hot-phonons in metallic single walled carbon nanotubes. ArXiv:cond-mat/0602366 (2006).
9. Kuroda, M.A., Leburton, J.-P.: Nonlinear transport and heat dissipation in metallic carbon nanotubes. Phys. Rev. Lett. **95**, 266803 (2005).
10. Kuroda, M.A., Leburton, J.-P.: Joule heating induced negative differential resistance in free-standing metallic carbon nanotubes. Appl. Phys. Lett. **89**, 103102 (2006).
11. Pop, E., Mann, D., Cao, J., Wang, Q., Goodson, K., Dai, H.: Negative differential conductance and hot phonons in suspended nanotube molecular wires. Phys. Rev. Lett. **95**, 155505 (2005).
12. Pop, E., Mann, D.A., Goodson, K.E., Dai, H.: Electrical and thermal transport in metallic single-wall carbon nanotubes on insulating substrates. J. Appl. Phys. **101**, 093710 (2007).
13. Javey, A., Guo, J., Paulsson, M., Wang, Q., Mann, D., Lundstrom, M., Dai, H.: High-field quasiballistic transport in short carbon nanotubes. Phys. Rev. Lett. **92**, 106804 (2004).
14. Green, F., Neilson, D.: Metallic carbon nanotubes as high-current-gain transistors. ArXiv:cond-mat/0602034 (2006).
15. Pozdnyakov, D.V., Galenchik, V.O., Komarov, F.F., Borzdov, V.M.: Electron transport in armchair single-wall carbon nanotubes. Physica E **33**, 336 (2006).
16. Pozdnyakov, D.V., Galenchik, V.O., Borzdov, V.M., Komarov, F.F.: Peculiar properties of electron transport in single-wall armchair carbon nanotubes. In: Proc. SPIE **6328**, pp. 0Y-1–9 (2006).

17. Park, J.-Y., Rosenblatt, S., Yaish, Y., Sazonova, V., Ustunel, H., Braig, S., Arias, T.A., Brouwer, P.W., McEuen, P.L.: Electron–phonon scattering in metallic single-walled carbon nanotubes. Nano Lett. **4**, 517 (2004).
18. Ragab, T., Basaran, C.: Joule heating in single-walled carbon nanotubes. J. Appl. Phys. **106**, 063705 (2009).
19. Jishi, R.A., Dresselhaus, M.S., Dresselhaus, G.: Electron-phonon coupling and the electrical coductivity of fullerene nanotubules. Phys. Rev. B **48**, 11385 (1993).
20. Gunlycke, D., Lawler, H.M., White, C.T.: Room-temperature ballistic transport in narrow graphene strips. Phys. Rev. B **75**, 085418 (2007).
21. Barisic, S., Labbe, J., Friedel, J.: Tight binding and transition-metal superconductivity. Phys. Rev. Lett. **25**, 919 (1970).
22. Pietronero, L., Strassler, S., Zeller, H.R., Rice, M.J.: Electrical conductivity of a graphite layer. Phys. Rev. B **22**, 904 (1980).
23. Lazzeri, M., Piscanec, S., Mauri, F., Ferrari, A.C., Robertson, J.: Electron transport and hot phonons in carbon nanotubes. Phys. Rev. Lett. **95**, 236802 (2005).
24. Piscanec, S., Lazzeri, M., Mauri, F., Ferrari, A.C., Robertson, J.: Kohn anomalies and electron–phonon interactions in graphite. Phys. Rev. Lett. **93**, 185503 (2004).
25. Chen, J.-H., Jang, Ch., Xiao, Sh., Ishigami, M., Fuhrer, M.S.: Intrinsic and extrinsic performance limits of graphene devices on $SiO_2$. Nature Nanotech. **3**, 206 (2008).
26. Lemay, S.G., Janssen, J.W., van den Hout, M., Mooij, M., Bronikowski, M.J., Willis, P.A., Smalley, R.E., Kouwenhoven, L.P., Dekker, C.: Two-dimensional imaging of electronic wavefunctions in carbon nanotubes. Nature **412**, 617 (2001).
27. Wirtz, L., Rubio, A.: The phonon dispersion of graphite revisited. Solid State Commun. **131**, 141 (2004).
28. Pozdnyakov, D., Galenchik, V., Borzdov, A., Borzdov, V., Komarov, F.: Influence of scattering processes on electron quantum states in nanowires. Nanoscale Res. Lett. **2**, 213 (2007).
29. Lim, Y.S., Yee, K.J., Kim, J.H., Shaver, J., Haroz, E.H., Kono, J., Doorn, S.K., Hauge, R.H., Smalley, R.E.: Coherent lattice vibrations in single-walled carbon nanotubes. Nano Lett. **6**, 2696 (2006).
30. Dubay, O., Kresse, G., Kuzmany, H.: Phonon softening in metallic nanotubes by a Peierls-like mechanism. Phys. Rev. Lett. **88**, 235506 (2002).
31. Pozdnyakov, D.V., Galenchik, V.O., Borzdov, A.V., Borzdov, V.M., Komarov, F.F.: Modeling of non-stationary electron–phonon transport in armchair single-wall carbon nanotubes. In: Phys., Chem. Appl. Nanostruct.: Rev. Short Notes to Nanomeeting–2007, pp. 245–248 (2007).
32. Purewal, M.S., Hong, B.H., Ravi, A., Chandra, B., Hone, J., Kim, Ph.: Scaling of resistance and electron mean free path of single-walled carbon nanotubes. Phys. Rev. Lett. **98**, 186808 (2007).
33. Matsuda, Y., Deng, W.-Q., Goddard, W.A.: Contact resistance properties between nanotubes and various metals from quantum mechanics. J. Phys. Chem. C **111**, 11113 (2007).
34. Yan, H., Song, D., Mak, K.F., Chatzakis, I., Maultzsch, J., Heinz, T.F.: Time-resolved Raman spectroscopy of optical phonons in graphite: Phonon anharmonic coupling and anomalous stiffening. Phys. Rev. B **80**, 121403 (2009).
35. Hale, P.J., Hornett, S.M., Moger, J., Horsell, D.W., Hendry, E.: Hot phonon decay in supported and suspended exfoliated graphene. Phys. Rev. B **83**, 121404 (2011).
36. Wang, H., Strait, J.H., George, P.A., Shivaraman, Sh., Shields, V.B., Chandrashekhar, M., Hwang, J., Rana, F., Spencer, M.G., Ruiz-Vargas, C.S., Park, J.: Ultrafast relaxation dynamics of hot optical phonons in graphene. Appl. Phys. Lett. **96**, 081917 (2010).
37. Bonini, N., Lazzeri, M., Marzari, N., Mauri, F.: Phonon anharmonicities in graphite and graphene. Phys. Rev. Lett. **99**, 176802 (2007).